\documentclass[twocolumn]{aastex61}

\newcommand\aastex{AAS\TeX}

\usepackage{mathtools}

\submitjournal{ApJ}

\shorttitle{\aastex\ Oversized gas clumps at extremely low metallicity}
\shortauthors{Shi et al.}
\begin{document}

\title{Over-sized gas clumps  in an extremely-metal-poor molecular cloud revealed by ALMA's pc-scale maps}

\correspondingauthor{Yong Shi}
\email{yshipku@gmail.com}

\author[0000-0002-8614-6275]{Yong Shi}
\affil{School of Astronomy and Space Science, Nanjing University, Nanjing 210093, China.}
\affil{Key Laboratory of Modern Astronomy and Astrophysics (Nanjing University), Ministry of Education, Nanjing 210093, China.}

\author{Junzhi Wang}
\affil{Shanghai Astronomical Observatory, Chinese Academy of Sciences, 80 Nandan Road, Shanghai 200030, China}

\author{Zhi-Yu Zhang}
\affil{School of Astronomy and Space Science, Nanjing University, Nanjing 210093, China.}
\affil{Key Laboratory of Modern Astronomy and Astrophysics (Nanjing University), Ministry of Education, Nanjing 210093, China.}

\author{Qizhou Zhang}
\affil{Center for Astrophysics $|$ Harvard \& Smithsonian, 60 Garden Street, Cambridge MA 02138, USA}

\author{Yu Gao}
\affil{Department of Astronomy, Xiamen University, Xiamen, Fujian 361005, China}
\affil{Purple Mountain Observatory, Chinese Academy of Sciences, Nanjing 210008, China}

\author{Luwenjia Zhou}
\affil{School of Astronomy and Space Science, Nanjing University, Nanjing 210093, China.}
\affil{Key Laboratory of Modern Astronomy and Astrophysics (Nanjing University), Ministry of Education, Nanjing 210093, China.}

\author{Qiusheng Gu}
\affil{School of Astronomy and Space Science, Nanjing University, Nanjing 210093, China.}
\affil{Key Laboratory of Modern Astronomy and Astrophysics (Nanjing University), Ministry of Education, Nanjing 210093, China.}

\author{Keping Qiu}
\affil{School of Astronomy and Space Science, Nanjing University, Nanjing 210093, China.}
\affil{Key Laboratory of Modern Astronomy and Astrophysics (Nanjing University), Ministry of Education, Nanjing 210093, China.}

\author{Xiao-Yang Xia}
\affil{Tianjin Astrophysics Center, Tianjin Normal University, Tianjin 300387}

\author{Cai-Na Hao}
\affil{Tianjin Astrophysics Center, Tianjin Normal University, Tianjin 300387}

\author{Yanmei Chen}
\affil{School of Astronomy and Space Science, Nanjing University, Nanjing 210093, China.}
\affil{Key Laboratory of Modern Astronomy and Astrophysics (Nanjing University), Ministry of Education, Nanjing 210093, China.}

%%\author{August Muench}
%%\affiliation{American Astronomical Society \\
%%2000 Florida Ave., NW, Suite 300 \\
%%Washington, DC 20009-1231, USA}
%%\collaboration{(AAS Journals Data Scientists collaboration)}

%% Note that the \and command from previous versions of AASTeX is now
%% depreciated in this version as it is no longer necessary. AASTeX 
%% automatically takes care of all commas and "and"s between authors names.

%% AASTeX 6.1 has the new \collaboration and \nocollaboration commands to
%% provide the collaboration status of a group of authors. These commands 
%% can be used either before or after the list of corresponding authors. The
%% argument for \collaboration is the collaboration identifier. Authors are
%% encouraged to surround collaboration identifiers with ()s. The 
%% \nocollaboration command takes no argument and exists to indicate that
%% the nearby authors are not part of surrounding collaborations.

%% Mark off the abstract in the ``abstract'' environment. 
\begin{abstract}

Metals are thought to have profound effects on the internal structures
of molecular clouds in which stars are born.  The absence of metals is
  expected  to prevent  gas  from  efficient cooling  and
fragmentation in theory.  However, this  effect  has not  yet been observed  at  low
metallicity  environments, such  as in  the early  Universe and  local
dwarf galaxies,  because of  the lack of  high  spatial resolution
maps of gas.  We carried out ALMA observations of the carbon monoxide (CO)
$J$=2-1 emission line at 1.4-parsec  resolutions of a molecular cloud
in  DDO 70  at 7\%  solar metallicity,  the  most metal-poor
galaxy currently known with a CO  detection.  In total, five clumps  have been identified
and they  are found to  follow more or  less the Larson's  law. Since
the CO  emission exists  in regions  with visual  extinction
$A_{\rm V}$ around 1.0, we converted  this $A_{\rm V}$ to the gas mass
surface density  using a  gas-to-dust ratio  of 4,594$\pm$2,848  for DDO
70.  We found  that the CO  clumps in  DDO 70 exhibit significantly
larger (on average  four times) sizes than those at  the same gas mass
surface densities in massive star-formation  regions of the Milky Way.
The  existence  of such  large  clumps appears to be consistent with
theoretical expectations that gas fragmentation in low metallicity clouds is suppressed. While our observation is  only for one cloud in
the galaxy,  if it  is representative, the  above result  implies
  suppressed gas fragmentation during the cloud collapse and star formation
 in  the early Universe.

\end{abstract}

%% Keywords should appear after the \end{abstract} command. 
%% See the online documentation for the full list of available subject
%% keywords and the rules for their use.


%% From the front matter, we move on to the body of the paper.
%% Sections are demarcated by \section and \subsection, respectively.
%% Observe the use of the LaTeX \label
%% command after the \subsection to give a symbolic KEY to the
%% subsection for cross-referencing in a \ref command.
%% You can use LaTeX's \ref and \label commands to keep track of
%% cross-references to sections, equations, tables, and figures.
%% That way, if you change the order of any elements, LaTeX will
%% automatically renumber them.

%% We recommend that authors also use the natbib \citep
%% and \citet commands to identify citations.  The citations are
%% tied to the reference list via symbolic KEYs. The KEY corresponds
%% to the KEY in the \bibitem in the reference list below. 

\section{Introduction} \label{sec:intro}

Across the cosmic  time, the rate of star formation  first rises until
redshift  of   $\sim$2  and  then   drops  toward  the   present  time
\citep{Madau14}.  The majority of cosmic star formation takes place in
galaxies with lower stellar masses  and metallicity than our Milky Way
\citep{Sobral14}, especially for star  formation in the early Universe
\citep{Knudsen16, Inoue16}.   Formation of  stars is a  local process,
taking place within clouds of molecular gas with sizes of only tens of
parcsecs (pcs). Metals (elements heavier  than helium) are expected to
have profound effects  on the internal structure of  a molecular cloud
and its collapse into stars \citep{Omukai05, Glover12}: metal poor gas
is  expected  to  be  warmer   due  to  inefficient  cooling  and  the
fragmentation  is  expected to  be  suppressed.   However, studies  of
nearby galaxies have not seen such an effect because of a lack of high
spatial-resolution   gas  maps   of   extremely  metal-poor   galaxies
\citep{Rubio15,     Schruba17}.     Using     the    Atacama     Large
Millimeter/sub-millimeter Array (ALMA), we  resolved a molecular cloud
in an extremely metal-poor  galaxy DDO 70 at a scale  of 1.4 pc.  This
resolution is  sufficient to  resolve the  Jeans fragmentation  with a
typical   length   of   $\sim$   3.5  pc   assuming   $T$=30   K   and
$n(H_{2})$=10$^{2}$ cm$^{-3}$ \citep{Zhang09}.

\begin{figure*}[tbh]
  \begin{center}
  \includegraphics[scale=0.7]{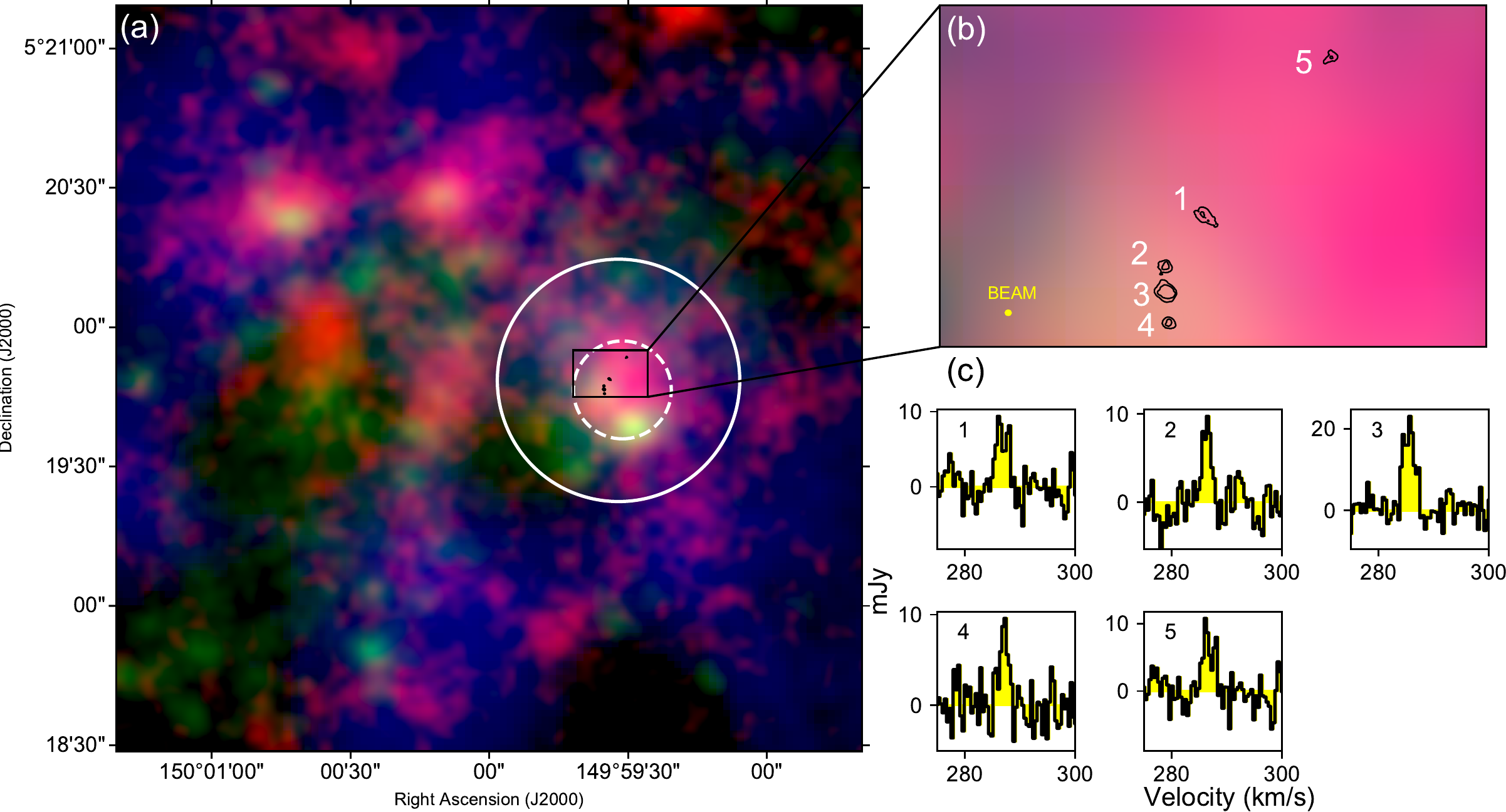}
  \caption{\label{img_spec}
  {\bf  False  color images  and  the  ALMA
      inteferometric data:} {\bf  a,} the false color image  of DDO 70
    where red is for infrared emission at 160 $\mu$m, green for the far-UV
    emission  and blue  for the  atomic hydrogen  21 cm  emission. The
    solid white circle  indicates the ALMA  FOV with the  white dashed circle
    for  the IRAM-30m  FOV. Black contours are for five detected CO clumps at levels of 1.0 (0.85-$\sigma$) and 4.5 mJy$\bullet$km/s/beam (3.84-$\sigma$). {\bf  b,} an  enlarged view  of
    detected CO clumps. {\bf  c,} the CO $J$=2-1 spectra of five
  detected CO clumps. }
\end{center}
\end{figure*}

DDO 70 is  a nearby dwarf irregular  galaxy at a distance  of 1.38 Mpc
\citep{Tully13}.   It   has  a  total  stellar   mass  of  $10^{7.56}$
$M_{\odot}$ \citep{Cook14} and a gas-phase  oxygen abundance of 7\% of
the  solar value  \citep{Kniazev05}.   Previously we  have targeted  a
star-forming region  in this galaxy as  shown in Figure~\ref{img_spec}
to  search  for   the  CO  $J$=2-1  emission  with   the  Institut  de
Radioastronomie  Millimetrique  (IRAM) 30-m  telescope  \citep{Shi16},
whose  detection  provides  direct   evidence  for  the  existence  of
molecular gas  at such a  low metallicity.   It is currently  the most
metal-poor  galaxy with  a CO  detection \citep{Elmegreen13,  Rubio15,
  Shi15, Schruba17}.  This single-dish detection offers an opportunity
to resolve the molecular cloud with ALMA.

\begin{figure}[tbh]
\begin{center}
 \includegraphics[scale=0.55]{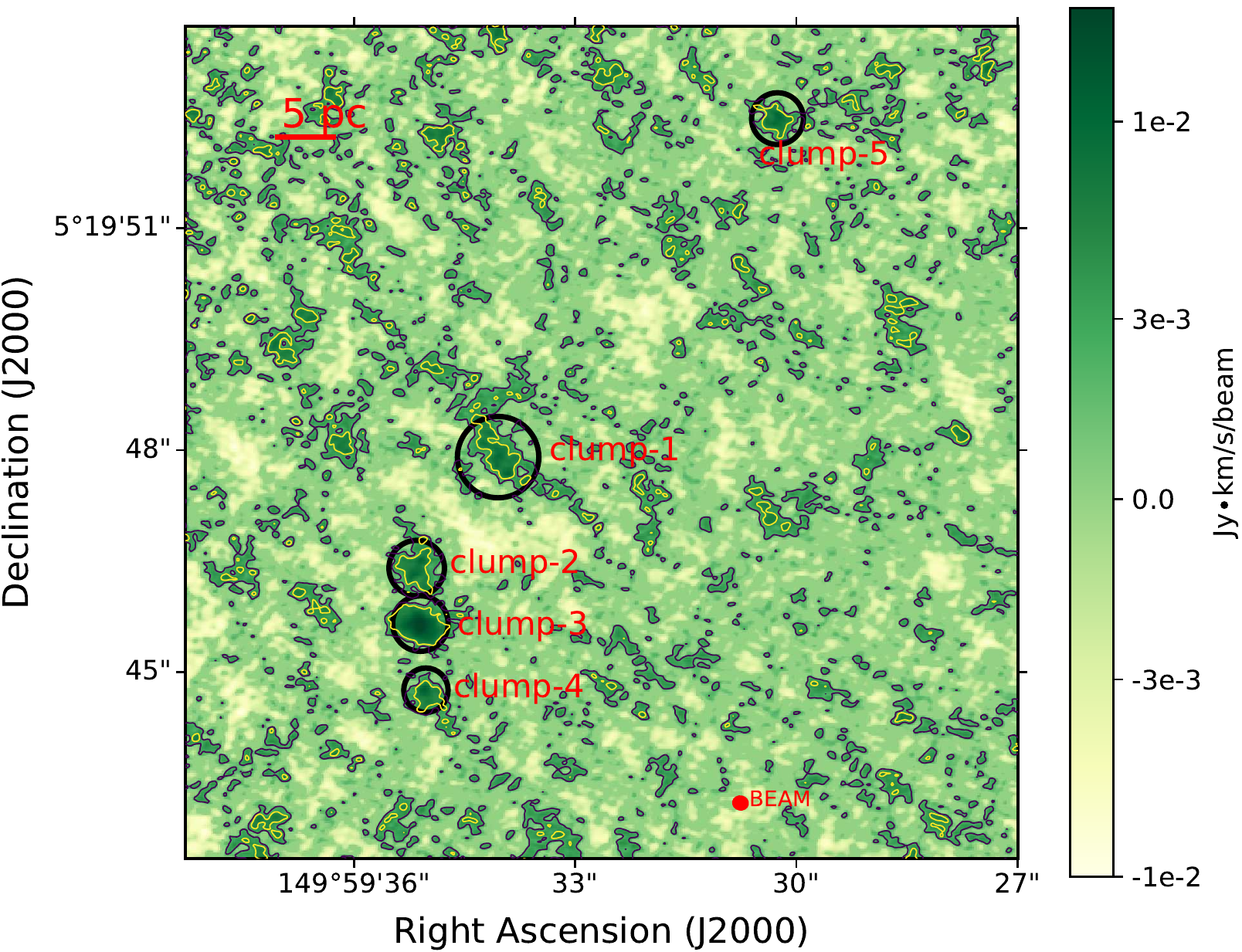}
 \caption{\label{ds9_fourcore} {\bf The ALMA CO $J$=2-1 images of five detected clumps in DDO 70.} The image is the moment zero integrated
   from 281 to 291 km/s. Each circle denotes a region that is used for imfit() in CASA to obtain the size of each clump (see text). The black and yellow contours stand for 1-$\sigma$ (2.3 mJy/beam$\bullet$km/s) and 2-$\sigma$ noise levels,
 respectively.}
\end{center}
\end{figure}

\section{Data} \label{sec:data}

\subsection{ALMA data of DDO 70}
To probe the internal structure of  the star-forming region in DDO 70,
we resolved  the CO $J$=2-1  emission using the  ALMA with a  field of
view of 26.4$''$ (177 pc) as shown in Figure~\ref{img_spec}.  The ALMA
observations (PID: 2016.1.00359.S,  PI: Y.  Shi) were  composed of two
array  configurations of  C40-4  and C40-7,  which  have an  angular
resolution  of  0.11  arcsec   and  0.39  arcsec,  respectively.  The
corresponding   integration  time   is  1.1   hours  and   0.6  hours,
respectively.  The observations were performed on 11 Nov.  2016 and 03
Aug 2017, respectively. We calibrated the data individually with CASA,
following  the standard  pipeline, then  combined all  visibility data
together for cleaning.  We selected  briggs weighting with a weighting
of 1.5,  to optimize  sensitivity and sidelobe  of the dirty beam
response.     A    final    datacube    with   a    beam    size    of
0.21$''$$\times$0.19$''$(1.47$\times$1.27   pc)  was   produced.   The
1-$\sigma$ sensitivity is about 0.9  mJy/beam at a velocity resolution
of 0.4  km/s.  We listed calibrators  in Table 1.  The  uncertainty of
the  absolute flux  calibration is  about 10\%  by comparing  with the
monitored fluxes of the ALMA calibrators.

To search  for CO  clumps, we  first identified   pixels that  have a
signal to noise  ratio (S/N) $>$ 3.5 in two  adjacent channels of ALMA
data. The search was done over the  velocity range of the IRAM 30-m CO
spectrum  (280-290 km/s).  For  regions with  more  than ten  detected
pixels ($>$ 15\% beam in area),  we  extracted  their  spectra   and  counted  those  with  an
integrated S/N $>$ 4 as detection.  A single Gaussian profile was used
to fit  the spectra. In  total we identified  five clumps as  shown in
Figure~\ref{ds9_fourcore} and listed in Table~\ref{tab_co_clumps}.  To
measure the size of  each CO clump, we used CASA imfit()  to fit a 2-D
Gaussian profile to the  velocity-integrated flux (moment-zero) map of
the    clump    within    a   circular   region.    As    shown    in
Figure~\ref{ds9_fourcore},  each  circle  is defined  to  be  slightly
larger than  the 1-$\sigma$ noise contour  (2.3 mJy/beam$\bullet$km/s)
to include  some background  emission but  not too  large so  that the
enclosed pixels are  dominated by noises. Note that the  circle is not
the  size of  the clump  but encloses  pixels over  which the  imfit()
performed the fitting. We tested  that, if the size of the circular
region increases
or  decreases by  20\%, the  changes in  the derived  clump sizes  are
negligible.  The  imfit() also returns the  associated uncertainties.
The derived  size dispersions  ($\sigma_{\rm ma}$,  $\sigma_{\rm mb}$)
along major  and minor axis  were used to  estimate the radius  of the
core:     $R$=$D$$\times$tan($\sqrt{\sigma_{\rm    ma}     \sigma_{\rm
    mb}}$)$\frac{3.4}{\pi^{0.5}}$ \citep{Solomon87}, where  $D$ is the
distance of DDO\,70. The defined radius represents the square root of
the  area of a clump.  The virial  mass of each core
was   calculated  using   $M_{\rm   vir}$  =   1040$R{\sigma_{v}^{2}}$
\citep{Solomon87},  where $\sigma_{v}$  is the  line width  dispersion
from the spectral fitting.

\subsection{Ancillary data of DDO 70}

The multi-wavelength  ancillary data of  DDO 70 were available  in the
literature  as  compiled  in  our  previous  work  \citep{Shi16}.  The
infrared data  at 70, 160, 250  and 350 $\mu$m were  retrieved from the
archive of the {\it Herschel}  Space Observatory and were reduced with
unimap \citep{Traficante11}.   We refer readers to our previous work
\citet{Shi14, Shi16} for details.
The  reduced {\it Spitzer} images  at  3.6, 4.5,  5.8 and  24
$\mu$m are from the Local Volume Legacy program \citep{Dale09}. The HI
image is from the program  of Local Irregulars That  Trace Luminosity
Extremes \citep{Hunter12}  and the  far-UV image  is available  in the
GALEX data archive (http:// galex.stsci.edu/GalexView/).

\section{The Larson's law}

As shown in Figure~\ref{img_spec}, in total five CO clumps are detected and all
of them are within the IRAM 30-m  coverage in spite of the larger ALMA
FOV.   As  listed  in   Table~\ref{tab_co_clumps},  the  total  fluxes
(127$\pm$11 mJy  km/s) of  these individual  clumps account  for about
80\% of  the emission (160$\pm$30 mJy  km/s) as seen by  the IRAM 30-m
telescope, indicating no significant missing fluxes.  These clumps are
resolved both spatially and spectroscopically  with radii of 1.5-3 pc,
velocity   dispersion   of   0.6-1   km/s   and   virial   masses   of
0.7-3$\times$10$^{3}$  $M_{\odot}$.  Their  CO luminosities  are about
20-60 K km/s pc$^{2}$, with  the corresponding kinematic CO conversion
factors   of    30-120   $M_{\odot}$   pc$^{-2}$    (K   km/s)$^{-1}$.

\begin{figure}[t]
  \begin{center}
    \includegraphics[scale=0.4]{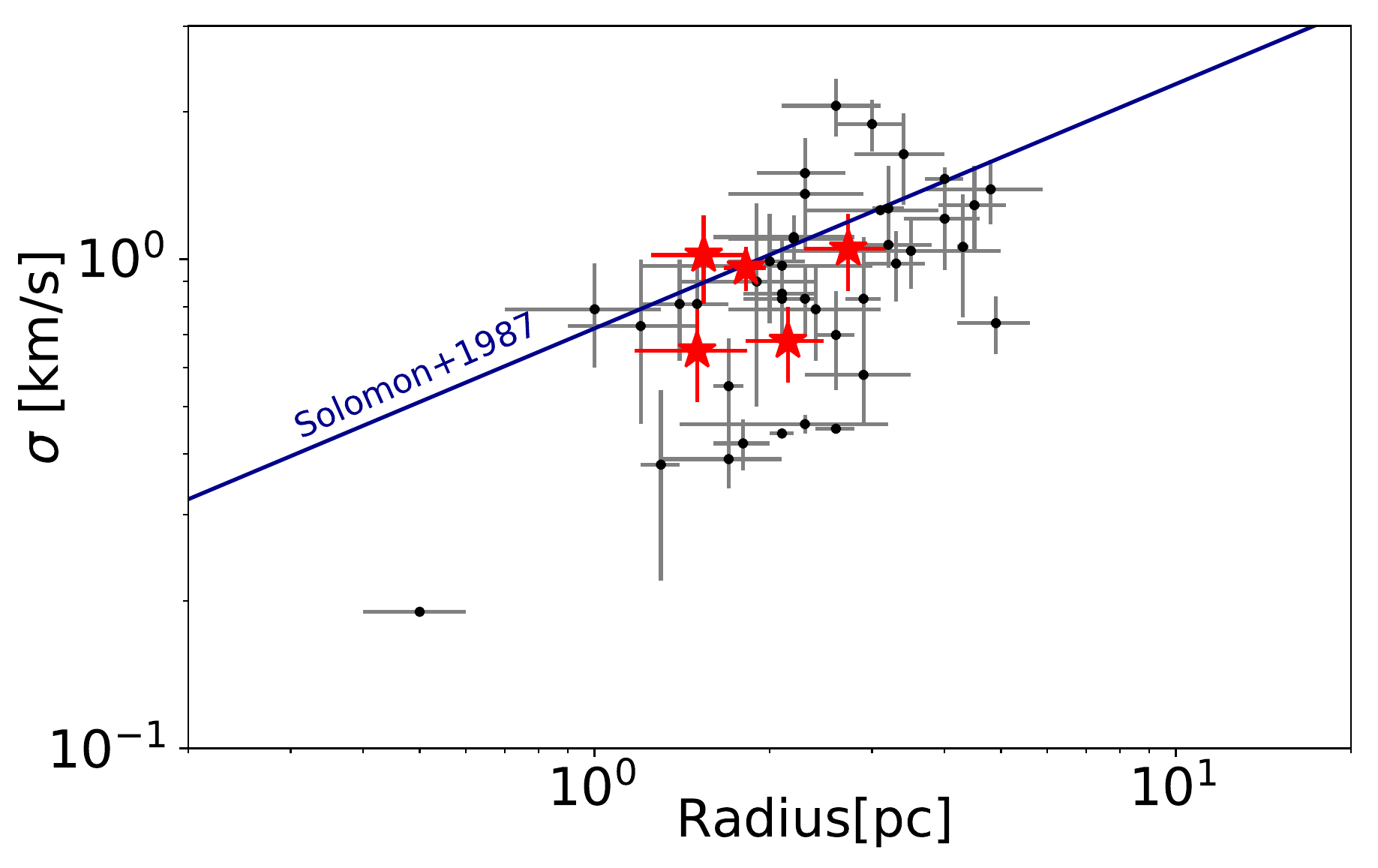}
  \caption{\label{larson_law}  {\bf CO  clumps  in DDO  70 follow  the
      Larson's  law of  the  line  width vs.   radius:}  The red  star
    symbols denotes five clumps detected in DDO 70, with black circles
    for CO clumps in the dwarf NGC 6822 \citep{Schruba17}.  The
    green   solid    line   is   the   best-fitted    one   of the  Milky
    Way \citep{Solomon87}. }
\end{center}
\end{figure}

Figure~\ref{larson_law} shows the distributions  of these CO clumps in
the so-called Larson's law \citep{Larson81, Solomon87, Schruba17} that
describes  the  relationship between  the  line  width and  the  clump
radius.  As  shown  in the  figure,  clumps  in  DDO  70 do  not  show
systematic deviations from those in the Milky Way and the dwarf galaxy
NGC 6822  at 20\%  solar metallicity.  Different  interpretations have
been proposed  for the origin  of Larson's law:  the clumps may  be in
virial  equilibrium, which  implies almost  constant gas  mass surface
densities ($\frac{M_{\rm  vir}}{{\pi}r^{2}}$) of CO clumps  from solar
to   7\%  solar   metallicity  \citep{Larson81,   Solomon87,  Rubio15,
  Schruba17}; the law could also  be an observational manifestation of
supersonic turbulence in which  the large-scale kinetic energy injects
into the ISM  and cascades to small scales  \citep{Kritsuk13}, and the
dynamics  of  clumps   at  pc-scales  are  thus   still  dominated  by
turbulence.    It   is   difficult    to   differentiate   these   two
interpretations   without   additional    data.    As   discussed   in
\citet{Kritsuk13},  the turbulence  origin could  result in  a flatter
slope of  the Larson's law  than the virial  origin, but our  few data
points do not have enough dynamic ranges to measure the slope.

Given that the  dust shielding drops with  the decreasing metallicity,
CO sizes should shrink. This is, however, in contrast to the fact that
CO  clumps  in DDO  70  have  similar sizes  to  CO  clumps at  higher
metallicity  as shown  in Figure~\ref{larson_law}.   We quantitatively
discuss this in the following section.

\section{Comparisons in the clump sizes between DDO 70 and the Milky Way}

\begin{figure}[t]
\begin{center}
 \includegraphics[scale=0.32]{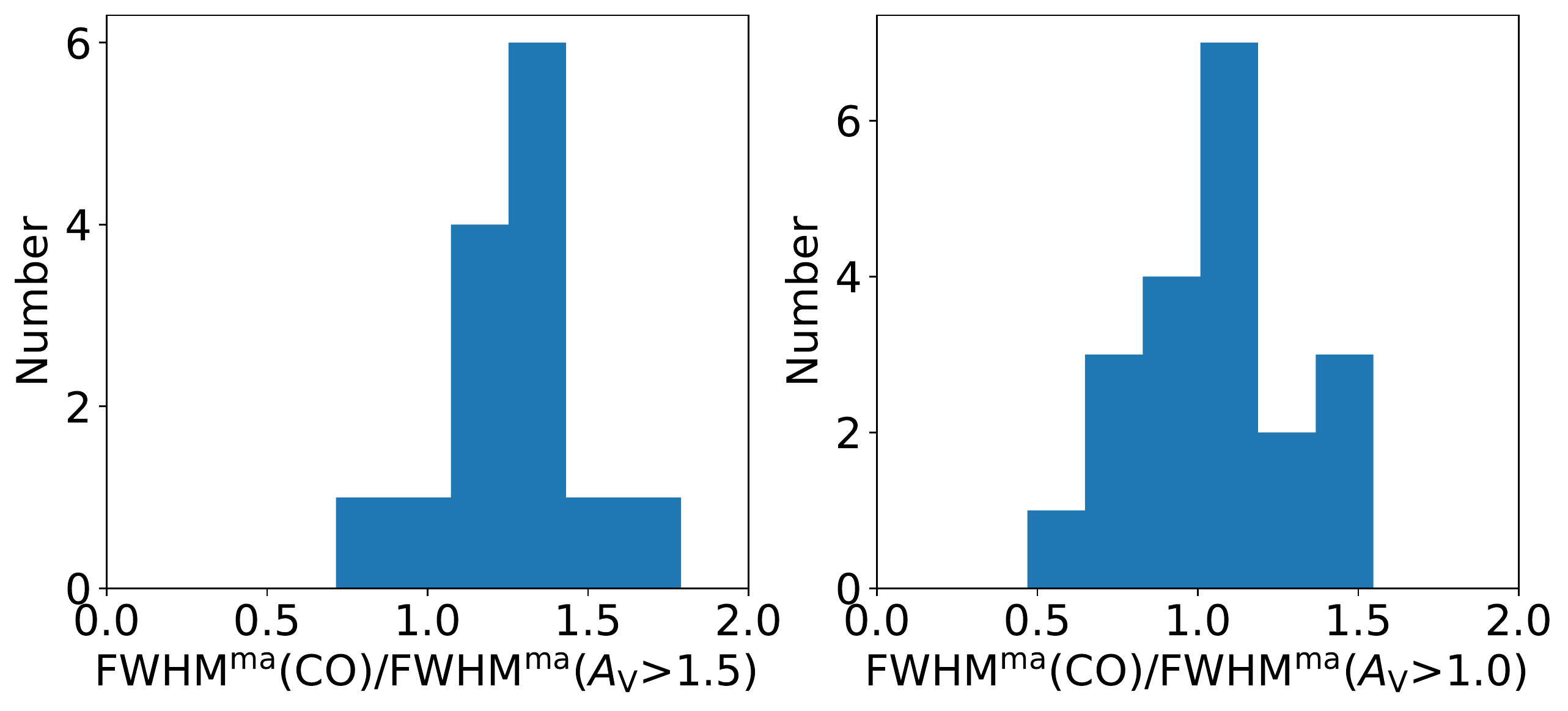}
 \caption{\label{comp_CO_dust} {\bf The ratio of CO clump sizes and dust clump sizes above two different extinction levels in the SMC.} }
\end{center}
\end{figure}

\subsection{The required shielding dust extinction for CO to exist in DDO 70}

In the Milky  Way, ultraviolet photons with energies  larger than 11.1
eV can photo-dissociate CO molecules,  so that CO emission only exists
in  the dust-shielded  regions with  a visual  extinction higher  than
1.0-2.0 mag \citep{Glover10, Bergin07}.  Further studies of metal-poor
dwarf galaxies  including Large Magellanic Cloud  and Small Magellanic
Cloud (SMC) confirm similar shielding  conditions for CO emission. And
the  relationship between  the  CO intensity  ($I_{\rm  CO}$) and  the
visual  extinction  ($A_{\rm  V}$)   shows  little  dependence  on  the
gas-phase  metallicity  \citep{Leroy09,   Lee15,  Lee18}.   Using  the
$I_{\rm CO}$-$A_{\rm V}$ relationship of the SMC \citep{Lee18}, our CO
clumps  are inferred  to  have  $A_{\rm V}$  between  1  and 2.5  mag,
supporting  the  above  required  shielding for  the  presence  of  CO
emission in DDO70.

To quantitatively  determine the extinction  levels above which dust clumps
have similar sizes to  CO clumps, we used the dust and  CO maps of the
SMC,  given  that the  galaxy  has  high-S/N  maps while  hosting  the
metallicity as close as possible to our target galaxy.  To produce the
dust map  of the SMC,  the {\it  Herschel} images were  retrieved from
Herschel  Inventory  of  the  Agents of  Galaxy  Evolution  (HERITAGE)
\citep{Meixner13}  at  100,  160,  250 and  350  $\mu$m  with  angular
resolutions of 7.7, 12, 18 and 25 arcsec, respectively.  Following the
previous work  \citep{Lee15}, we  used the single  modified black-body
(MBB) model  with a fixed dust  emissivity of 1.5 to  fit the Herschel
SED to produce the dust temperature maps at 350$\mu$m resolutions.  We
then interpolated  this map to  the 160$\mu$m resolution  and produced
the                extinction                 map                using
$\tau_{160{\mu}m}$=$I_{160{\mu}m}/B_{\nu}(T_{\rm   dust},  160{\mu}m)$
and $A_{\rm V}$=2,200$\tau_{160{\mu}m}$  \citep{Lee15}.  The CO $J$=2-1
data with an  angular resolution of 28 arcsec were  retrieved from the
archive of the  APEX telescope and reduced in a  standard way.  The CO
clump sizes were measured in the same  way as for the CO data in DDO 70.  Sizes
of dust clumps  were measured above two $A_{\rm V}$  values of 1.0 and
1.5, respectively:  we first subtracted  constant $A_{\rm V}$  (1.0 or
1.5)  from the  extinction map  of  the SMC  and convolved  to the  CO
resolution, and then measured  the sizes in the same way  as for the CO data
in DDO 70. Using spatially-resolved clumps that are defined to have sizes
larger than  1.2 times the FWHM,  we calculated the ratio  of the 
major-axis  FWHM of CO and the FWHM of the dust emission.   In total,
we had  14 clumps  with
$A_{\rm  V}$  $>$  1.5  and  20   clumps  with  $A_{\rm  V}$  $>$  1.0,
respectively.  As shown in Figure~\ref{comp_CO_dust}, the median ratio
with  a  standard  deviation   is  1.29$\pm$0.24  for  CO/dust($A_{\rm
  V}$$>$1.5) and  1.05$\pm$0.26 for CO/dust($A_{\rm V}$$>$1.0).   As a
result, the dust clumps with $A_{\rm V}$ $>$ 1.0 or $\tau_{160{\mu}m}$ $>$
1/2,200 have similar sizes to  CO clumps. Since $\tau_{160{\mu}m}$ is a
direct  measurement  from  the  MBB  fitting  and  thus  suffers  less
systematic uncertainties as compared to $A_{\rm V}$, we will adopt the
limiting $\tau_{160{\mu}m}$=1/2,200 in the following analysis. Although
the SMC  maps have poorer spatial  resolutions than the maps  of our DDO
70, the  relationship between $I_{\rm  CO}$ and extinction  has little
dependence on  the spatial resolution down  to sub-pc as done  for the
Milky Way \citep{Lee15, Lee18}, implying  that the above result of the
SMC likely holds at pc  scales of our maps. While the
  $I_{\rm CO}$-$A_{\rm V}$ relationship shows little dependence on the
  metallicity from  the Milky Way to  the SMC at 20\%$Z_{\odot}$, we here  do assume that
  such an independence can extend to DDO 70 at 7\%$Z_{\odot}$.

\begin{figure}[t]
\begin{center}
 \includegraphics[scale=0.42]{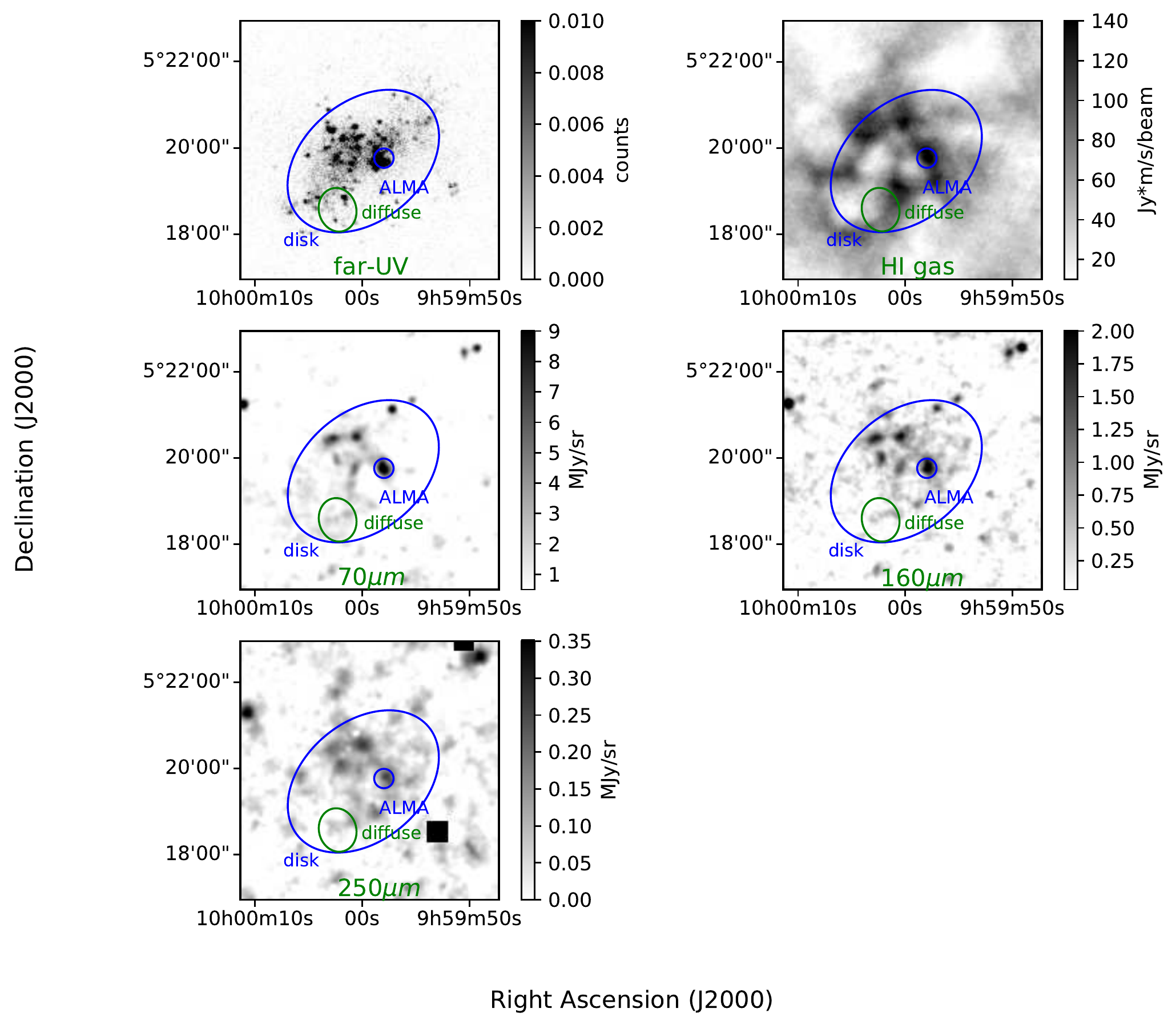}
 \caption{\label{all_imgs} {\bf Multi-wavelength images of DDO70}. The large blue ellipse encircles
 the whole disk of the galaxy. The small blue circle indicates the ALMA FOV and the green ellipse denotes
 the diffuse region. }
\end{center}
\end{figure}

\begin{figure}[t]
\begin{center}
 \includegraphics[scale=0.31]{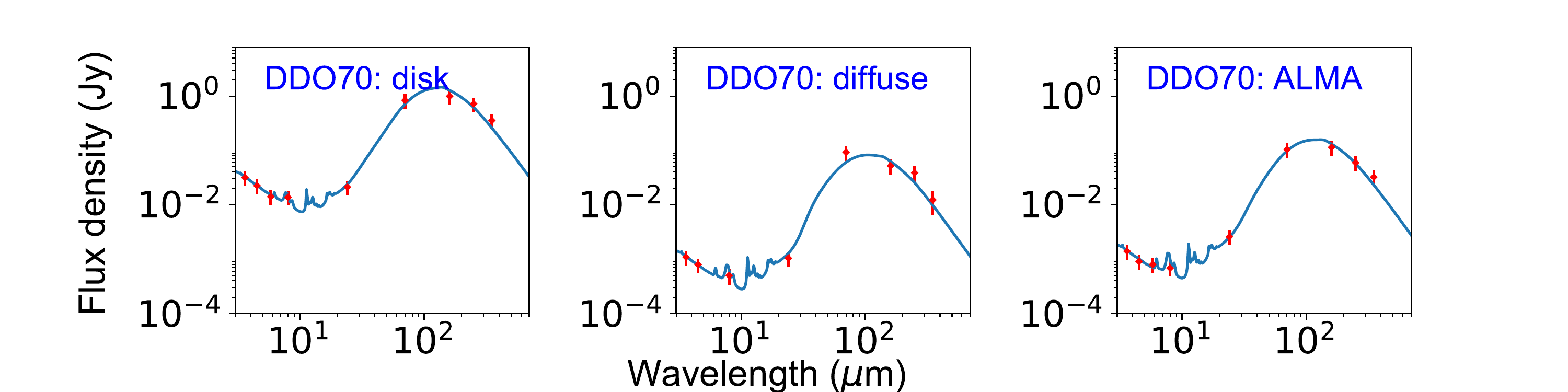}
 \caption{\label{sed_m1_MW} {\bf Infrared SEDs of the whole disk, the diffuse region and the ALMA aperture in DDO 70.}
 Red diamonds are the {\it Spitzer} and {\it Herschel} photometry with 1-$\sigma$ error
bar that includes additional 30\% systematic flux uncertainties. The blue line is the best-fitted dust model \citep{Draine07}. }
\end{center}
\end{figure}

\subsection{The gas-to-dust (GDR) ratio of DDO 70} 
  
The required  extinction for CO to  exist in the Milky  Way ($A_{\rm
    V}$=1.0  mag) corresponds  to a  gas  mass surface  density of  14
  M$_{\odot}$/pc$^{2}$        assuming        $N_{\rm        H}/A_{\rm
    v}$=1.87$\times$10$^{21}$  cm$^{-2}$ mag$^{-1}$  \citep{Draine03}.
  At  lower metallicities,  the required  extinction corresponds  to a
  higher  gas  surface density  given  the  corresponding larger  GDR,
  making CO  suitable to trace  gas of a  high surface density  at low
  metallicity  \citep{Rubio15, Schruba17}.  Arguably,
    one of the  most reliable methods to measure the  GDR at extremely
    low metallicity  is to use  the ratio of  the atomic gas  and the
    dust  mass in  the  diffuse region  where the  total  cold gas  is
    dominated   by   the  atomic   gas   instead   of the molecular   gas
    \citep{Israel97, Dame01, Bolatto11, Shi14, Jameson16}.

Following   the   methods   of  our   previous   work
  \citep{Shi14}, we first carried  out the photometric measurements of a diffuse region
  as well as  the whole disk  of DDO 70 and the region within
  the  ALMA  aperture.   These  regions  are
  illustrated  in  Figure~\ref{all_imgs}  for   far-UV,  HI  gas,  and
  infrared bands at 70, 160 and 250 $\mu$m, respectively.  The diffuse
  region  was  identified  as  extended  emission  at all {\it  Herschel}
  wavelengths, not overlaid with  star-formation regions, and was used
  to infer the gas-to-dust ratio (GDR) of the galaxy. In these regions
  the total cold gas  is dominated by atomic gas, so  the ratio of the
  atomic  gas   to  the  dust   mass  can   be  treated  as   the  GDR
  \citep{Leroy11, Sandstrom13, Shi14}.  The  results of {\it Herschel}
  and   {\it   Spitzer}   infrared  photometry   were   presented   in
  Table~\ref{tab_ir_photometry}.  Similar  to what have been  done for
  other  extremely  metal-poor  galaxies \citep{Shi14,  Fisher14},  we
  fitted   the  full   infrared  SEDs   with  the   DL07  dust   model
  \citep{Draine07}   to  estimate   the  dust   masses  as   shown  in
  Figure~\ref{sed_m1_MW} and  listed in Table~\ref{tab_fit}.  The GDR
  of  the galaxy  is thus  derived to  be $6078^{+1000}_{-1600}$, referred as the diffuse-region GDR.   If
  adopting  SMC or  LMC dust  grains, the  derived ratio  increases by
  about 10-20\%.  This  ratio roughly corresponds to the  trend of the
  GDR $\propto$ $Z^{-\alpha}$ with $\alpha$=1.6 given the Galactic GDR
  of 100.  An  $\alpha$ larger than unity is consistent  with the fact
  that  the   dust-to-metal  ratio   decreases  with   the  decreasing
  metallicity as  measured in absorption-line  systems \citep{DeCia13}
  as  well as  the  global GDR  of  dwarf galaxies  where  the HI  gas
  dominates   the   total   cold   gas   at   global   galaxy   scales
  \citep{Remy-Ruyer14}.

 From the above measurements, we  can infer other physical properties
  of the region  within the ALMA FOV.  It contains a dust  mass of 530
  $M_{\odot}$,  a total  gas mass  of 3.2$\times$10$^{6}$  $M_{\odot}$
  given   the  above diffuse-region   GDR   and    a   molecular   gas   mass   of
  2.7$\times$10$^{6}$  $M_{\odot}$  after  subtracting an  HI  mass  of
  0.53$\times$10$^{6}$  $M_{\odot}$.   It  has   a  stellar   mass  of
  1.7$\times$10$^{6}$  M$_{\odot}$  and  a   star  formation  rate  of
  1.6$\times$10$^{-4}$   $M_{\odot}$/yr  from   the  3.6   $\mu$m  and
  24$\mu$m+far-UV   photometry,    respectively,   as    detailed   in
  \cite{Shi14}.

 \subsection{The required shielding gas surface density for CO to exist in DDO 70 and its associated uncertainty}

As  argued above,  the size  of the  CO clump  is set  by the  spatial
extent above a  threshold extinction ($\tau_{160{\mu}m}$=1/2,200) for
CO to survive. This can be  converted to the required gas mass surface
density  with  $\tau_{\lambda}$=$\kappa_{\lambda}$$\Sigma_{\rm  dust}$
and the corresponding  GDR of DDO 70, where  $\kappa_{\lambda}$ is the
mass absorption  coefficient of  dust and  $\Sigma_{\rm dust}$  is the
dust  mass surface  density.  We derived  the  required shielding  gas
density and its uncertainty with the following steps:

(1) The threshold $\tau_{160{\mu}m}$ above  which the dust  clump size
represents  the  CO clump  size:  as  discussed above,  this  limiting
$\tau_{160{\mu}m}$ is 1/2,200, and above  this extinction the size of a
dust clump  is similar to  the CO size  within 26\%.  We  thus adopted
26\% as the error of the limiting $\tau_{160{\mu}m}$.

(2) The diffuse-region GDR of DDO 70:  the diffuse-region GDR of DDO  70 is derived to be  $6078^{+1000}_{-1600}$,  where  the  error
reflects    the   photon    noise    plus    30\%   systematic    flux
uncertainties. Additional 15\% uncertainty is included for the case if
adopting different dust  models (SMC or LMC) as stated  above. The GDR
across  a  galaxy is  roughly  constant  if removing  the  metallicity
gradient,  with  variation  up  to 50\%  based  on  spatially-resolved
studies  \citep{Shi14, Draine14,  Sandstrom13}.  By  adding the  above
three   errors  quadratically,   we  have   a  final   uncertainty  of
56\%. Note that our  derived GDR is near  the lower
  bound at similar metallicities  in the study of \citet{Remy-Ruyer14}
  if   assuming  the   CO  conversion   factor  increases   with  the decreasing
  metallicity, implying an even larger shielding gas surface density.

(3) The  difference in the  MBB fitting  vs.  DL07 dust  modeling: the
  extinction  map  of  the  SMC  used to  derive  the  above  limiting
  $\tau_{160{\mu}m}$=1/2,200 is  based on the  MBB fitting to  the {\it
    Herschel} SED.   The GDR of DDO  70 is, however, derived  from the
  fitting to the {\it Spitzer}+{\it  Herschel} SED using the DL07 dust
  model.  This  could cause some  systematic offset.  As a  result, we
  added {\it Spitzer} photometry to the SED of the SMC and carried out
  DL07  dust  modeling.   By  adopting  $\kappa_{\rm  160{\mu}m}$=13.1
  cm$^{2}$/g  \citep{Weingartner01, Li01},  we obtained  the DL08/MBB  dust
  ratio of 1.89$\pm$0.37.   The above GDR of DDO 70  is thus corrected
  to be 11487 with a 1-$\sigma$ error of 59\%.

(4) The drop in  the GDR from the diffuse region  to the dense region:
  the above  diffuse-region GDR is derived  from the
  diffuse region. Studies suggest a decrease  in the GDR by a factor
  of  2-3 from  the  diffuse to  the  dense region  in  the Milky  Way
  \citep{Jenkins09,    Planck11},    SMC   and    LMC    \citep{Bot04,
    Roman-Duval14}.  It seems  that such a decrease is  not a function
  of the galaxy metallicity.  We  thus adopted a factor of 2.5$\pm$0.5
  to  correct the  above  GDR  to be  4,594  with a 1-$\sigma$ error  of
  62\%. We are aware of  the work by \citet{Roman-Duval17} that claims
  a factor  of 7 drop in  the GDR of the  SMC from the diffuse  to the
  dense region.  However,  their diffuse  region has a  very low  gas density
  mainly located in the out-skirts of  the SMC, resulting in a very high GDR
  of  1.5$\times$10$^{4}$ that  is a  factor of  10 larger  than other
  works  \citep{Bouchet85, Gordon09,  Leroy11, Roman-Duval14}.  Such a
  region is  different from our  diffuse region which is  still within
  the  galaxy disk, and our diffuse region has $\Sigma_{\rm HI}$=11.6 M$_{\odot}$/pc$^{2}$ as compared to that of the ALMA region around 21 M$_{\odot}$/pc$^{2}$.  We further  used their  result to  show that  the
  adopted drop  with a  factor of 2.5$\pm$0.5  is approximate  for our
  study. In DDO 70, our defined diffuse region has a dust mass surface
  density of 1.9x10$^{-3}$ M$_{\odot}$/pc$^{2}$ that is a factor of 10
  lower than that of the ALMA  region.  Their equations 13 and 14 give
  the relationships between dust mass surface densities and total cold
  gas mass surface densities for  LMC and SMC, respectively. These two
  equations indicate that the GDR drops  by a factor of 2.1-3.2 if the
  dust mass  surface density increases  by a  factor of 10  over their
  observed dynamic ranges.

By adopting the above final GDR (4594$\pm$2848), $\tau_{160{\mu}m}$=1/2,200
and  $\kappa_{\rm 160{\mu}m}$=13.1  cm$^{2}$/g, we  derived the  gas
mass surface  densities of  CO clumps  in DDO  70 to  be 756$\pm$468
M$_{\odot}$/pc$^{2}$.

\begin{figure*}[tbh]
  \begin{center}
    \includegraphics[scale=0.7]{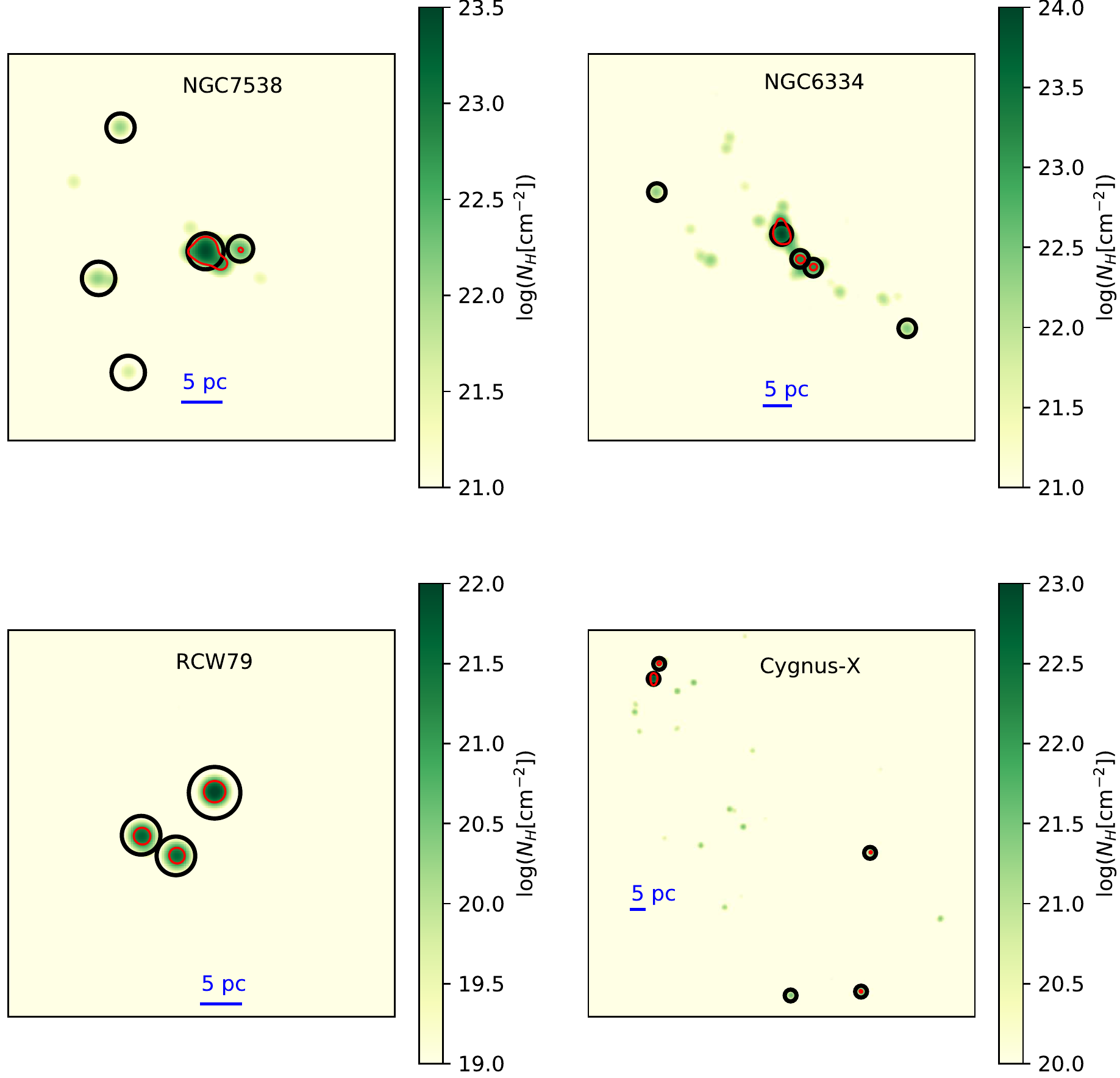}
\caption{\label{MW_regions} {\bf Maps of dust clumps with $\Sigma_{\rm
    gas}^{\rm limit}$=700 M$_{\odot}$/pc$^{2}$  in four massive star
  formation regions  of the  Milky Way.} The  $\Sigma_{\rm gas}^{\rm
  limit}$ is the  contour level above which a clump  is defined (see
text). All  images have  the same physical  pixel sizes  and spatial
resolutions as the CO map of DDO  70. In each panel, the red contour
represents  the noise  level  of the  CO  map of  DDO  70 where  the
1-$\sigma$  noise  is  10\%  of  the peak  intensity  of  the  whole
map. Clumps  without red  contours are small
  ones  below the  contour level.   Black circles  are the  defined
regions  for  imfit()  to  fit  the profiles  to  obtain  the  clump
sizes. All circles are larger than the mimicked noise contour. }   
\end{center}
\end{figure*}

\begin{figure*}[tbh]
  \begin{center}
    \includegraphics[scale=0.7]{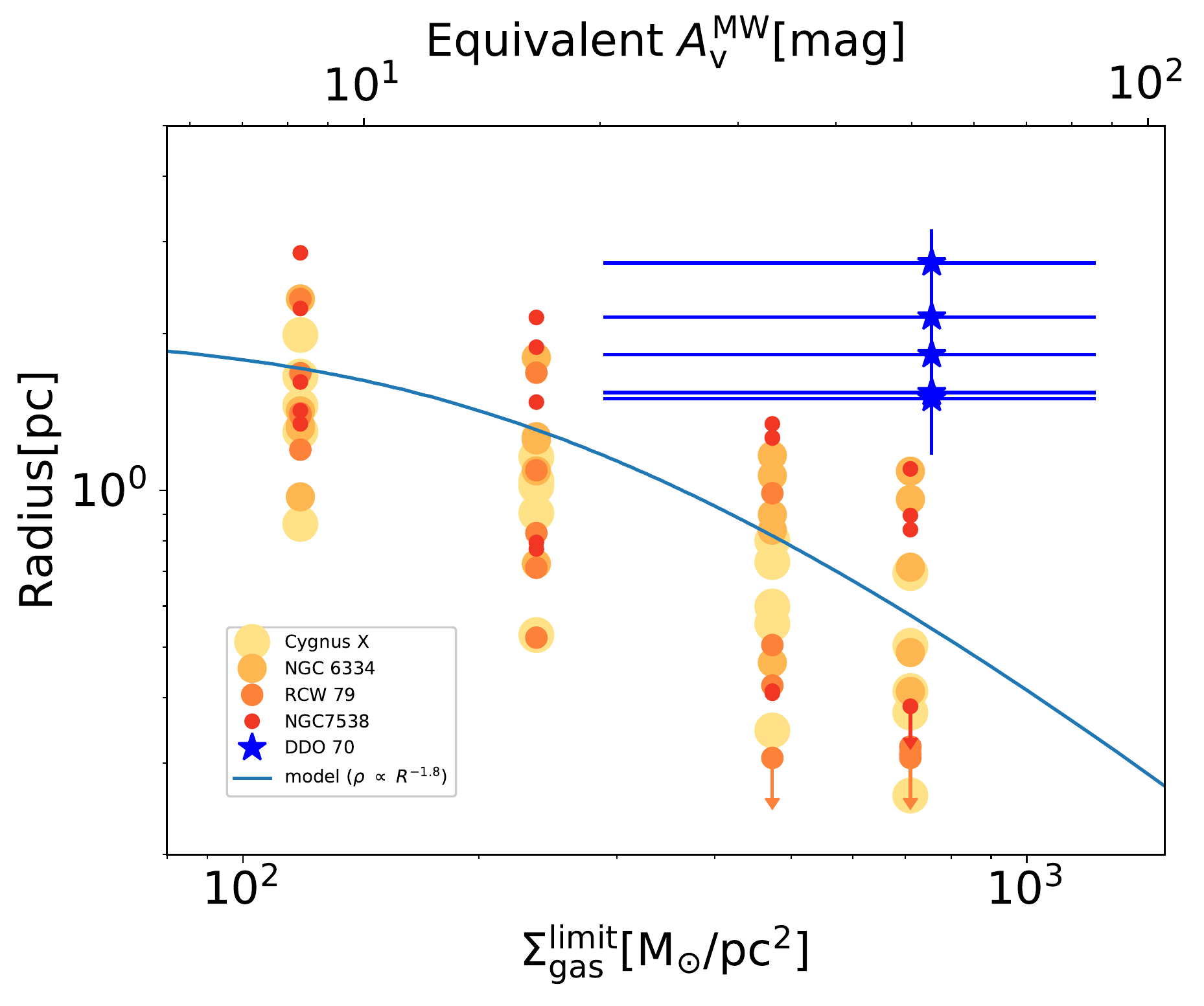}
  \caption{\label{Av_size} {\bf CO clumps  in DDO 70 are significantly
      larger than dust clumps in massive star-formation regions of the
      Milky Way.}   The $\Sigma_{\rm gas}^{\rm limit}$  is the contour
    level above which  a clump is defined.  The solid  line is a clump
    with   a   density   profile   of   $R^{-1.8}$   \citep{Mueller02,
      Beuther02}. Note that,  because we only detected  five clumps in
    DDO  70, we  thus  compared to  the five  largest  clumps in  each
    massive star-formation region of the Milky Way. }
\end{center}
\end{figure*}

\subsection{Sizes of clumps in massive star-formation regions of the Milky Way}

In  order  to  evaluate  whether gas  clumps  at  low
  metallicity show some systematic differences from those at the solar
  metallicity, we  compared clumps  in DDO  70 to  those in  the Milky
  Way.  We  measured the clump sizes  in the Milky Way  using maps of
gas  mass  surface  densities  of four  massive  star-forming  regions
including Cygnus  X \citep{Cao19}, NGC 6344  \citep{Russeil13}, RCW 79
\citep{Liu17} and  NGC 7538 \citep{Fallscheer13} at  distances of 1.4,
1.75,        4.3        and       2.7        kpc,        respectively.
 Table~\ref{global_property}   compares  some   global
  properties of these  regions to our targeted region in  DDO 70. The
  total cold gas mass is  derived from the {\it Herschel} dust emission as
  shown below, and the SFR is estimated from the  WISE 22 $\mu$m by
  assuming $f_{22{\mu}m}$=$f_{24{\mu}m}$  \citep{Leroy08}. As listed
  in the  table, these  regions have similar  global gas  mass surface
  densities to that of DDO 70.  But their global SFR surface densities
  are significantly  larger that that  in DDO 70, consistent  with our
  previous study  that the SFR  efficiency is eliminated  at extremely
  low metallicity \citep{Shi14}.

The gas maps in the Milky Way are based on the
dust  maps through  the MBB  fitting to  the {\it  Herschel} SED  with
$\beta$=2,  $\kappa_{\rm  300{\mu}m}$=10  cm$^{2}$/g and  GDR  of  100
\citep{Beckwith90}.  We further corrected  these maps with two factors
in order  to have fair comparisons  with DDO 70.  The  first factor is
about the dust  model.  As shown in the above  section, the dust model
for  DDO 70  and SMC  is the  silicate-graphite-PAHs interstellar  one
\citep{Weingartner01, Li01},  which gives $\kappa_{\rm  300{\mu}m}$ of
4.81 cm$^{2}$/g. As a result, we  increased the original densities by a
factor of 2.1. Another factor is the  drop in the GDR from the diffuse
region to  the dense region  as also  discussed in the  above section,
which  makes the  gas densities  decrease  by a  factor of  2.5. We therefore
corrected the map  by multiplying  with a final  factor of
0.83. To  remove the effects  of different spatial  resolutions and
pixel sizes on the clump size measurements, we further re-binned these
maps to  the  same physical pixel  sizes as DDO  70's CO  map and
convolved with  Gaussian functions to  the same  physical spatial
resolutions as DDO  70's.

Similar to what has been done for the measurements of dust clump sizes
in SMC, we  first subtracted a series of given  surface densities from
the  whole maps  of  the  above regions,  and  only  used pixels  with
positive values.  We denoted  these surface densities  as $\Sigma_{\rm
  gas}^{\rm limit}$,  which represent  the contour levels  above which
clumps are  defined.  We then  used the same method  as for DDO  70 to
identify clumps  and measured  the deconvolved  clump radii  that were
obtained by quadratically subtracting the  beam size from the measured
one. In the DDO 70's CO map, the 1-$\sigma$ noise is about 10\% of the
peak intensity of its whole map. To  remove the effect of the noise in
the derived clump sizes, when defining a circle for fitting a clump in
the Milky Way, we make sure that the  size of the circle is larger than the
DDO 70's percentage noise level that  is 10\% of the peak intensity of
the whole  map.  An  example is  shown in  Figure~\ref{MW_regions} for
four  Milky  Way's  regions with  $\Sigma_{\rm  gas}^{\rm  limit}$=700
M$_{\odot}$/pc$^{2}$, and the corresponding  clump sizes are listed in
Table~\ref{clump_MW}.  As shown  in the  figure, some
  clumps are  below the  DDO 70's  percentage noise  level, indicating
 that they  would not  have been  detected in  DDO 70.  Such non-detection
  implies that  the Milky Way  should have on average  smaller
  clump  sizes. Since  we  have five  detected clumps  in  DDO 70,  we
  compare them to the five largest  clumps in each region of the Milky
  Way including those below the  DDO 70's percentage noise level.  If
the derived size is smaller than  the {\it Herschel} resolution of the
map, we used it as the upper-limit.

\subsection{RESULTS}

In Figure~\ref{Av_size},  the sizes of  our CO  clumps in DDO  70 were
compared to those in the Milky Way at different $\Sigma_{\rm gas}^{\rm
  limit}$.  Since we only detected five clumps in DDO 70, we thus only
used the five largest ones in  each Milky Way's region for comparison.
Note  that  $\Sigma_{\rm gas}^{\rm  limit}$  is  not the  average  gas
density of a clump within its  size, but the contour level above which
the region is defined. In the  figure, we also overlaid a modeled trend
for a clump with a mass of 4000  $M_{\odot}$, a radius of 2.0 pc and a
density profile of $\propto$ $R^{-1.8}$.   This density profile is the
mean of star-forming  clumps found in the  Milky Way \citep{Mueller02,  Beuther02}.

As shown in  the figure, for the same  $\Sigma_{\rm gas}^{\rm limit}$,
the CO clumps in DDO70 show much  larger sizes than those in the Milky
Way.  The mean radius of the former  is 2.0 pc, while the latter has a
mean of $\sim$0.5 pc including upper-limits.   All clumps in DDO 70 are
larger than  1.5 pc  while all  clumps in  the Milky  Way at  the same
$\Sigma_{\rm  gas}^{\rm limit}$  are below  1.5  pc.  Even  at the  gas
density of  288 M$_{\odot}$/pc$^{2}$ (1-$\sigma$ lower  bound), DDO 70
has on average two times larger clump sizes than all four star-forming
regions in the Milky Way. In the  Herschel Infrared GALactic plane survey,  a sample of
$\sim$ 1.5$\times$10$^{4}$  dense clumps  has been identified  and all
their radii are smaller than 1.3 pc \citep{Veneziani17}.  This further
excludes the  statistical outliers in the  Milky Way
as the  cause of the  large clumps in  DDO 70.  In  principle, similar
analyses can  be carried out for  other dwarf galaxies such  as WLM at
13\%$Z_{\odot}$   \citep{Elmegreen13,  Rubio15}   and   NGC  6822   at
20\%$Z_{\odot}$ \citep{Schruba17}.   But their relatively  smaller GDR
as compared  to DDO 70  will make the  result inconclusive due  to the
large error  bars of the final  derived gas densities as  shown in the
Figure~\ref{Av_size}.

The  existence  of large  CO  clumps  in  DDO  70 is  consistent  with
theoretical  expectation of suppressed gas fragmentation  at
extremely low metallicity.  The high gas temperature results in a high
sound   speed,   low   Mach    number   and   thus   weak   turbulence
\citep{Glover12}.  Therefore large clumps cannot further fragment into
small ones \citep{Omukai05}. High gas temperature may
  also cause top-heavy stellar initial  mass functions as seen both in
  the  metal-poor LMC  and high-z  starburst galaxies  \citep{Zhang18,
    Schneider18}. The  difficulty in  gas fragmentation  at extremely
low metallicity suggests a small number of dense clumps in which stars
can  form.   This offers  the  physical  origin for  inefficient  star
formation at  low metallicity \citep{Shi14, Filho16,  Shi18}, implying
the suppression  of gas  fragmentation and  subsequent collapse  in the
early  Universe. Unfortunately,  our result  is  not
  statistically important, given that our  observation is only for one
  cloud. It  is difficult to  justify whether  it is typical  for all
  metal-poor clouds with the current data-set.

\section{Conclusions}\label{sec:conclusion}

In order to probe the internal  structure of a molecular cloud with an
extremely low metallicity, we carried out ALMA high-spatial-resolution
(1.4 pc)  observations of  CO emission  in DDO  70, currently  the most
metal-poor galaxy  with CO detection.  Five clumps in  total have been
detected and  account for  the majority  of the  single-dish emission,
indicating  that  the CO  emission  is  mostly confined  into  compact
regions at  low metallicity.   The size  and velocity  measurements of
these clumps suggest that they follow more or less the Larson's law of
the Milky Way.  A further comparison in sizes of  these clumps with
those in the  Milky Way indicates that they  have systematically larger
sizes, which should not be  caused by statistical outliers.  The large
sizes  may  be  the  result  of suppressed  gas  fragmentation  at  low
metallicity  as large  clumps  cannot further  fragment into  small
ones due to large Jeans mass and weak turbulence. As our
 observation is only for one region in DDO 70, it is crucial to carry out
 more studies to justify whether our result is representative of all types of
 metal-poor galaxies.

\acknowledgments

We thank the anonymous referee  for his/her important and constructive
suggestions that improve the quality of the paper significantly.  Y.S.
acknowledges the support  from the National Key R\&D  Program of China
(No.   2018YFA0404502,  No.   2017YFA0402704),  the  National  Natural
Science  Foundation  of  China  (NSFC grants  11825302,  11733002  and
11773013), and the Tencent Foundation through the XPLORER PRIZE.  J.W.
thanks  the  support  of  NSFC  (grant  11590783).  YG's  research  is
supported by  National Key Basic  Research and Development  Program of
China (grant No. 2017YFA0402704),  National Natural Science Foundation
of China (grant Nos. 11861131007, 11420101002), and Chinese Academy of
Sciences   Key   Research   Program  of   Frontier   Sciences   (grant
No.  QYZDJSSW-SLH008).  This  paper makes  use of  the following  ALMA
data: ADS/JAO.ALMA$\#$2016.1.00359.S.   ALMA is  a partnership  of ESO
(representing its member states), NSF (USA) and NINS (Japan), together
with NRC  (Canada) and NSC  and ASIAA  (Taiwan) and KASI  (Republic of
Korea), in  cooperation with  the Republic of  Chile.  The  Joint ALMA
Observatory is operated by ESO, AUI/NRAO and NAOJ.

\clearpage
%------------------------------------------------

\begin{table*}[tbh]
\begin{center}
\caption{\label{tab_alma_info} ALMA observational information.}
\begin{tabular}{lllllll}\hline
SB name                           & DDO70$\_$a$\_$06$\_$TM2    &   DDO70$\_$a$\_$06$\_$TM1  \\
Array configuration                & C40-4                   &    C40-7                \\
\hline
Observing Date                    & 11-Nov-2016             & 03-Aug-2017             \\
Bandpass Calibrator               & J1058+0133              & J1058+0133            \\
Flux Calibrator                   & J0854+2006              & J1037-2934              \\
Gain Calibrator                   & J1008+0621              & J1008+0621            \\
Integration Time (s)              & 2177                    & 3810                  \\
Median PWV (mm)                   & 0.55                    & 0.7                      \\
Angular Resolution ($''$)         & 0.35                    & 0.10                    \\
\hline
\end{tabular}
\tablecomments{PWV stands for precipitable water vapour}
\end{center}
\end{table*}

%------------------------------------------------

\begin{table*}[b]
\begin{center}
\caption{\label{tab_co_clumps} Properties of CO $J$=2-1 clumps in DDO 70.}
\begin{tabular}{llllllllllllll}
\hline
region    & RA(J2000) & DEC(J2000) &  FWHM$_{\rm ma}$  &  FWHM$_{\rm mb}$  &  Radius  &  $\sigma_{v}$   & $S_{\rm CO}$${\Delta}V$    & $M_{\rm vir}$  & $L_{\rm CO}$    \\
          &           &            &  (pc)            &  (pc)            &  (pc)   &  (km/s)            &  (mJy km/s)           & (M$_{\odot}$)  & (K km/s pc$^{2}$) \\
\hline
clump-1 & 9:59:58.270 & +5:19:47.88 & 5.23$\pm$1.07 & 2.14$\pm$0.54 & 2.73$\pm$0.44 & 1.05$\pm$0.19 & 22.3$\pm$5.2 & 3102.1$\pm$1239.4 & 25.9$\pm$6.0\\
clump-2 & 9:59:58.344 & +5.19.46.37 & 2.75$\pm$0.60 & 2.55$\pm$0.54 & 2.15$\pm$0.33 & 0.68$\pm$0.12 & 16.4$\pm$3.3 & 1050.1$\pm$398.1 & 19.1$\pm$3.8\\
clump-3 & 9:59:58.343 & +5:19:45.62 & 2.88$\pm$0.33 & 1.74$\pm$0.20 & 1.82$\pm$0.15 & 0.96$\pm$0.10 & 51.1$\pm$6.8 & 1751.7$\pm$398.3 & 59.4$\pm$7.9\\
clump-4 & 9:59:58.336 & +5:19:44.72 & 1.88$\pm$0.54 & 1.81$\pm$0.60 & 1.50$\pm$0.33 & 0.65$\pm$0.14 & 16.1$\pm$4.0 & 662.0$\pm$319.8 & 18.7$\pm$4.7\\
clump-5 & 9:59:58.019 & +5:19:52.42 & 2.55$\pm$0.60 & 1.41$\pm$0.40 & 1.54$\pm$0.29 & 1.02$\pm$0.21 & 21.8$\pm$5.4 & 1679.9$\pm$763.4 & 25.4$\pm$6.3\\

\hline
\end{tabular}
\end{center}
\end{table*}

%------------------------------------------------

\begin{table*}
\tiny
\begin{center}
\caption{\label{tab_ir_photometry} Infrared photometry of the whole disk, diffuse region and ALMA aperture of DDO 70.}
\begin{tabular}{llllllllllllll}
\hline
region & Right ascension & Declination & sizes(ma,mb) & f(3.6$\mu$m) & f(4.5$\mu$m) & f(5.8$\mu$m) & f(8.0$\mu$m) & f(24$\mu$m) & f(70$\mu$m) & f(160$\mu$m) & f(250$\mu$m) & f(350$\mu$m) \\
       & (J2000)         & (J2000)     & (arcsec)     & (mJy)        & (mJy)        & (mJy)        & (mJy)        & (mJy)       & (mJy)       & (mJy)        & (mJy)       & (mJy)   \\ 
\hline 
disk    & 09 59 59.9 & +05 19 42 &118.5x82.7 &31.74$\pm$0.03 &22.51$\pm$0.04 &14.19$\pm$0.16 &13.82$\pm$0.15 &21.37$\pm$0.47 & 839$\pm$16 & 992$\pm$17 & 720$\pm$ 23 & 357$\pm$ 17 \\
diffuse & 10 00 02.3 & +05 18 34 &30.6x26.0  &1.09$\pm$0.01 &0.79$\pm$0.01 & $<$0.19 &0.50$\pm$0.06 &1.04$\pm$0.11 & 92$\pm$3 & 52$\pm$4 & 38$\pm$5 & 13$\pm$4 \\
ALMA    & 09 59 58.0 & +05 19 46 &13.5x13.5 &1.41$\pm$0.01 &0.91$\pm$0.01 &0.80$\pm$0.03 &0.69$\pm$0.03 &2.58$\pm$0.05 & 105$\pm$1 & 114$\pm$2 & 59$\pm$3 & 32$\pm$3 \\
\hline
\end{tabular}
\end{center}
\end{table*}

%---------------

\begin{table*}
\begin{center}
\caption{\label{tab_fit} The fitting results of the infrared SEDs.}
\begin{tabular}{llllllllllllll}
\hline
region & U$_{\rm min}$ & U$_{\rm max}$(fixed) & $\gamma$ & $\chi^{2}$/dof & $M_{\rm dust}$ & $M_{\rm HI}$/$M_{\rm dust}$ \\
 & & & & & (M$_{\odot}$) & \\
\hline

DDO70/disk           &    3.0   & 10$^{6}$ &     0.00 &     1.95 & (7.2$^{+2.4}_{-1.5}$)x10$^{3}$      & (2.2$^{+0.4}_{-0.7}$)x10$^{3}$     \\
DDO70/diff-1         &    7.0   & 10$^{6}$ &     0.00 &     3.06 & (2.1$^{+0.6}_{-0.4}$)x10$^{2}$      & (6.1$^{+0.4}_{-0.7}$)x10$^{3}$      \\
DDO70/ALMA           &    5.0   & 10$^{6}$ &     0.00 &     1.08 & (5.3$^{+1.3}_{-1.0}$)x10$^{2}$      & (1.$^{+0.4}_{-0.7}$)x10$^{3}$      \\

\hline
\end{tabular}
\tablecomments{In
addition to the flux uncertainties in  Table~\ref{tab_ir_photometry},
absolute calibration errors (20\%) are included when performing the fits.}
\end{center}
\end{table*}

%---------------

\begin{table*}
\begin{center}
\caption{\label{global_property} Global properties of star-formation regions in DDO 70 and the Milky Way.}
\begin{tabular}{llllllllllllll}
\hline
region &  D             & area                            & $M_{\rm gas}$           &  $\Sigma_{\rm gas}$    &  SFR              &     $\Sigma_{\rm SFR}$  \\ 
       & (kpc)          & (pc$^{2}$)                      &  (M$_{\odot}$)          & (M$_{\odot}$/pc$^{2}$) &  (M$_{\odot}$/yr)   &   (M$_{\odot}$/yr/kpc$^{2}$)  \\
\hline
DDO 70    (ALMA FOV)        &       1.38$\times$10$^{3}$  &     2.46$\times10^{4}$ &  3.2$\times10^{6}$    &    130              & 1.6$\times10^{-4}$     & 6.4$\times10^{-3}$ \\
NGC 6334  (Herschel FOV)    &       1.75                 &      2.30$\times10^{3}$ &   4.1$\times10^{5}$  &     179             & 5.0$\times10^{-4}$     & 0.22    \\
Cygnus-X  (Herschel FOV)    &       1.40                 &      1.44$\times10^{4}$ &   1.6$\times10^{6}$  &     109             & 1.4$\times10^{-3}$     & 0.10     \\
RCW 79    (Herschel FOV)    &       4.30                 &      1.50$\times10^{3}$ &   1.8$\times10^{5}$  &     122             & 2.1$\times10^{-4}$     & 0.14    \\
NGC 7538  (Herschel FOV)    &       2.70                 &      2.14$\times10^{3}$ &   3.3$\times10^{5}$  &     156             & 1.1$\times10^{-4}$     & 0.05     \\
\hline
\end{tabular}
\tablecomments{All associated errors are dominated by the systematic uncertainties. The gas mass of DDO 70 has an error around 30\%, while those
  in the Milky Way is around 10\%. All SFRs have errors around 20\%.  }
\end{center}
\end{table*}

\begin{table*}
\begin{center}
\caption{\label{clump_MW} Sizes of dust clumps with $\Sigma_{\rm gas}^{\rm limit}$=700 M$_{\odot}$/pc$^{2}$  in  massive star-formation regions of the Milky Way.}
\begin{tabular}{llllllllllllll}
\hline
Region     &  FWHM$_{\rm ma}$  &  FWHM$_{\rm mb}$  &  Radius   \\ 
           &   (pc)           &   (pc)           & (pc)      \\
\hline
Cygnus-X   & 2.04$\pm$0.08   & 0.36$\pm$0.15   & 0.69$\pm$0.14    \\
Cygnus-X   & 0.58$\pm$0.10   & 0.36$\pm$0.16   & 0.37$\pm$0.09    \\
Cygnus-X   & 0.41$\pm$0.00   & 0.25$\pm$0.00   & 0.26$\pm$0.00    \\
Cygnus-X   & 0.95$\pm$0.03   & 0.27$\pm$0.09   & 0.41$\pm$0.07    \\
Cygnus-X   & 1.03$\pm$0.10   & 0.37$\pm$0.25   & 0.50$\pm$0.18    \\
NGC6334    & 1.84$\pm$0.09   & 0.96$\pm$0.07   & 1.09$\pm$0.05    \\
NGC6334    & 1.43$\pm$0.27   & 0.97$\pm$0.29   & 0.96$\pm$0.17    \\
NGC6334    & 1.16$\pm$0.31   & 0.66$\pm$0.32   & 0.71$\pm$0.20    \\
NGC6334    & 0.92$\pm$0.03   & 0.27$\pm$0.09   & 0.41$\pm$0.07    \\
NGC6334    & 0.89$\pm$0.03   & 0.40$\pm$0.04   & 0.49$\pm$0.03    \\
RCW79      & 0.40$\pm$0.00   & 0.37$\pm$0.00   & 0.31$\pm$0.00    \\
RCW79      & 0.50$\pm$0.03   & 0.31$\pm$0.04   & 0.32$\pm$0.02    \\
RCW79      & 0.50$\pm$0.00   & 0.31$\pm$0.01   & 0.32$\pm$0.00    \\
NGC7538    & 1.58$\pm$0.12   & 1.23$\pm$0.10   & 1.13$\pm$0.06    \\
NGC7538    & 1.41$\pm$0.22   & 0.75$\pm$0.26   & 0.84$\pm$0.16    \\
NGC7538    & 0.43$\pm$0.00   & 0.39$\pm$0.00   & 0.33$\pm$0.00    \\
NGC7538    & 2.39$\pm$0.25   & 0.50$\pm$0.38   & 0.89$\pm$0.34    \\
NGC7538    & 0.40$\pm$0.00   & 0.32$\pm$0.00   & 0.29$\pm$0.00    \\

\hline
\end{tabular}
\end{center}
\end{table*}

%% This command is needed to show the entire author+affilation list when
%% the collaboration and author truncation commands are used.  It has to
%% go at the end of the manuscript.
%\allauthors

%% Include this line if you are using the \added, \replaced, \deleted
%% commands to see a summary list of all changes at the end of the article.
%\listofchanges

\end{document}